\documentclass[floatfix,prl,twocolumn,showpacs,superscriptaddress]{revtex4-1}

\usepackage{graphicx}
\usepackage{amsmath}
\usepackage{amsthm,amsmath,amsfonts,dsfont}
\usepackage{color}

\newcommand{\comment}[1]{}

\begin{document}

\title{Multicritical behavior in dissipative Ising models}

\author{Vincent R. Overbeck}
\affiliation{Institut f\"ur Theoretische Physik, Leibniz Universit\"at Hannover,
  Appelstra{\ss}e 2, 30167 Hannover, Germany}
\email[]{vincent.overbeck@itp.uni-hannover.de}

\author{Mohammad F. Maghrebi}
\affiliation{Joint Quantum Institute and Joint Center for Quantum
Information and Computer Science, NIST/University of Maryland, College
Park, Maryland 20742, USA}
\author{Alexey V. Gorshkov}
\affiliation{Joint Quantum Institute and Joint Center for Quantum
Information and Computer Science, NIST/University of Maryland, College
Park, Maryland 20742, USA}
\author{Hendrik Weimer}
\affiliation{Institut f\"ur Theoretische Physik, Leibniz Universit\"at Hannover,
  Appelstra{\ss}e 2, 30167 Hannover, Germany}

\begin{abstract}

We analyze theoretically the many-body dynamics of a dissipative Ising
model in a transverse field using a variational approach. We find that
the steady state phase diagram is substantially modified compared to
its equilibrium counterpart, including the appearance of a
multicritical point belonging to a different universality
class. Building on our variational analysis, we establish a
field-theoretical treatment corresponding to a dissipative variant of a
Ginzburg-Landau theory, which allows us to compute the upper critical
dimension of the system. Finally, we present a possible experimental
realization of the dissipative Ising model using ultracold Rydberg
gases.

\end{abstract}


\maketitle

The continuous transition between a paramagnetic and a ferromagnetic
phase within the Ising model in a transverse field is one of the most
important examples of a quantum phase transition. At finite
temperature, thermal fluctuations dominate while the phase transition
between the two phases remains continuous \cite{Sachdev1999}. Here, we
show that adding dissipation to the model strongly modifies the phase
diagram and gives rise to a multicritical point belonging to a
different universality class.

Rapid experimental progress in the control of tailored dissipation
channels
\cite{Syassen2008,Baumann2010,Barreiro2011,Krauter2011,Barontini2013},
combined with prospects to use dissipation for the preparation of
interesting many-body states
\cite{Diehl2008,Verstraete2009,Weimer2010}, has put dissipative
quantum many-body systems at the forefront of ultracold atomic
physics, quantum optics, and solid state physics. In particular,
systems driven to highly excited Rydberg atoms have emerged as one of
the most promising routes
\cite{Raitzsch2009,Carr2013,Malossi2014,Schempp2014,Urvoy2015,Weber2015,Goldschmidt2016,Lee2011,Honer2011,Glatzle2012,Ates2012a,Lemeshko2013a,Hu2013,Honing2013,Otterbach2014,Sanders2014,Hoening2014,Marcuzzi2014,Marcuzzi2016},
as the dissipation and interaction properties of Rydberg gases can be
very widely tuned \cite{Low2012}. These crucial experimental advances
have led to the investigation of driven-dissipative models in a wide
range of theoretical works
\cite{Glatzle2012,Goldstein2015,Goldschmidt2016,Hu2013,Sieberer2013,Wilson2016,LeBoite2013,Joshi2013,Tomadin2010,Tomadin2011,Jin2014,Marino2016,Torre2010,Mascarenhas2015,Cui2015}.
However, the theoretical understanding of dissipative quantum
many-body systems is still in its infancy, as many of the concepts and
methods from equilibrium many-body systems cannot be applied. As a
consequence, little is known even about the most basic dissipative
models.

\begin{figure}[t]
\centering
\includegraphics[width=\linewidth]{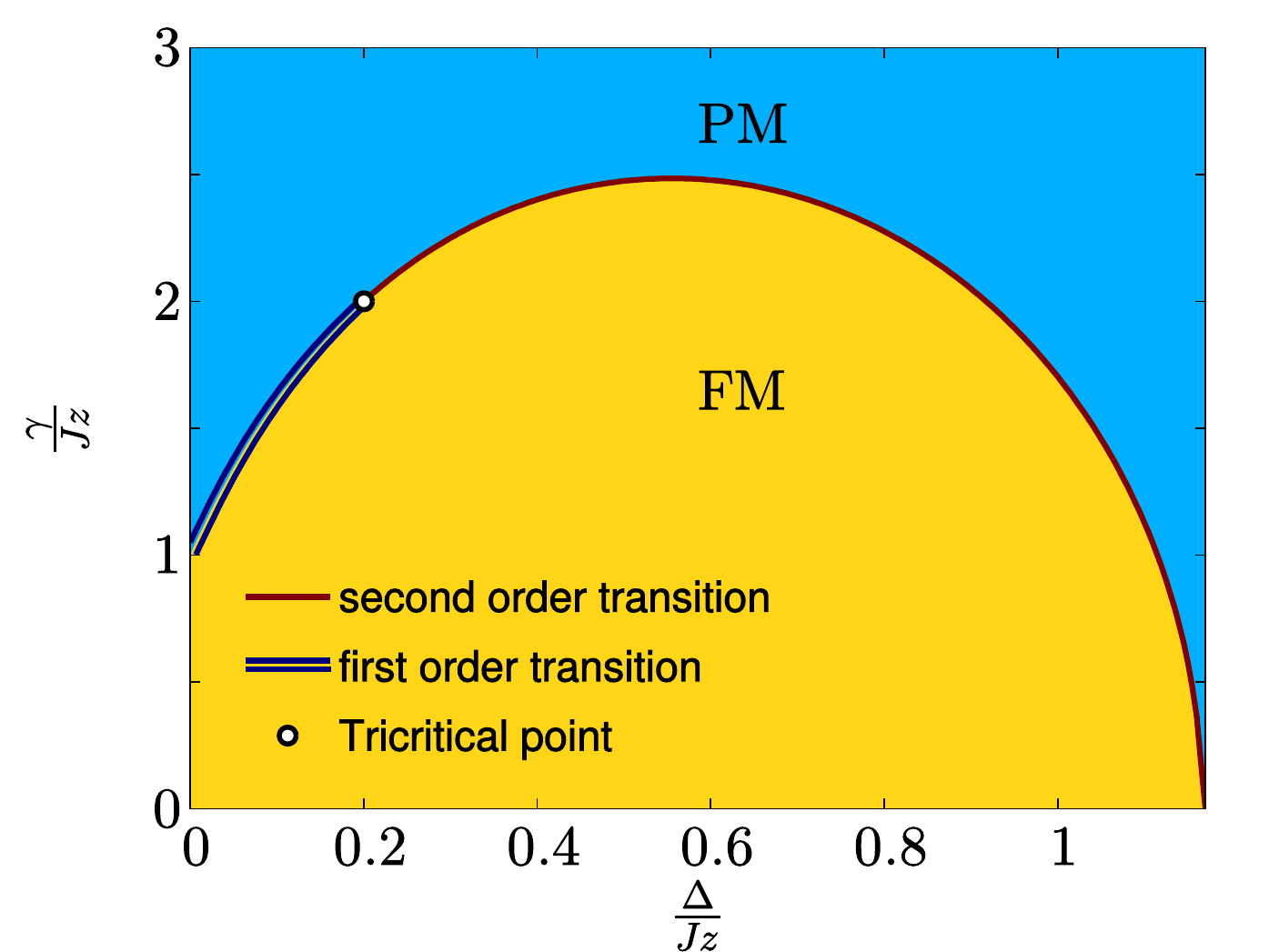}

\caption{Phase diagram of the three-dimensional dissipative Ising
  model according to the variational principle based on product states. The system can undergo phase transitions between
  ferromagnetic (FM) and paramagnetic (PM) phases, which can be either
  continuous or first order. The continuous and first order transition
  lines meet at a tricritical point.}

\label{fig:phasediagram}
\end{figure}

In this Letter, we perform a variational analysis of the steady state
of dissipative Ising models using a recently introduced variational
method. In contrast to the equilibrium case, we find that the
continuous transition is replaced by a first order transition if the
dissipation is sufficiently stronger than the transverse field, see
Fig.~1. Strikingly, we find that the model gives rise to a multicritical
behavior, as the two types of transitions are connected by a
tricritical point. This deviation from the equilibrium situation
underlines the fact that dissipative many-body systems constitute an
independent class of dynamical systems that go beyond the presence of
a finite effective temperature. Furthermore, we establish a
field-theoretical treatment of dissipative many-body systems
corresponding to a Ginzburg-Landau theory, which allows us to identify
the upper critical dimension of the tricritical point. Finally, we
give a concrete example of a possible experimental realization of
dissipative Ising models based on Rydberg-dressed atoms in optical
lattices, showing that the observation of the tricritical point is
within reach in present experimental setups.

Dissipative systems are no longer governed by the unitary
Schr\"odinger equation, but have to be described in terms of a quantum
master equation instead. Here, we consider the case of a Markovian master equation for the density operator $\rho$, given in the Lindblad form as
\begin{align}
  \frac{d}{dt}\rho=-i[H,\rho]+ \sum\limits_{i} \left(c_i\rho
    c_i^{\dagger}-\frac{1}{2}\{c_i^\dagger c_i,\rho\} \right).
 \label{eq:mastereq}
\end{align}
Importantly, dissipative quantum systems generically relax towards one
or more steady states, which can be found by solving the equation
$d \rho /dt  = 0$. 

For the dissipative Ising model, the Hamiltonian is of the form
 \begin{align}
  H={\Delta}\sum_{i}\sigma_z^{(i)}-J\sum_{\langle i j \rangle}\sigma_x^{(i)}\sigma_x^{(j)}
  \label{eq:Hamiltonian},
\end{align}
where $\Delta$ denotes the strength of the transverse field and $J$
indicates the strength of the ferromagnetic Ising interaction.  The
quantum jump operators $c_i = \sqrt{\gamma} \sigma_-^{(i)}$ describe
dissipative spin flips occuring with a rate $\gamma$.  Consequently,
this dissipative Ising model is a straightforward generalization
including Lindblad dynamics. Note that the present model
  is unrelated to a series of similarly named models, where a strong
  coupling to the bath is present \cite{Werner2004,Werner2005} or
  where explicit time-dependent driving is considered
  \cite{Goldstein2015}. We would also like to stress that in contrast
to previous studies of dissipative Rydberg gases
\cite{Lee2011,Ates2012a,Hu2013,Hoening2014,Weimer2015,Weimer2015a,Overbeck2016},
the present model exhibits a global $Z_2$ symmetry. Since the
dissipation acts in the eigenbasis of the transverse field, the master
equation is invariant under applying a $\sigma_z$ transformation to
all the spins. Different driven-dissipative models with
  $Z_2$ symmetry have been investigated previously
  \cite{Lee2013,Torre2013}. Crucially, this $Z_2$ symmetry can be
spontaneously broken by the steady state of the dynamics, constituting
a continuous dissipative phase transition. Interestingly, recent
results obtained within the Keldysh formalism show that this
continuous transition can break down for sufficiently strong
dissipation \cite{Maghrebi2016}, hinting that the dissipative phase
diagram is much richer than its equilibrium counterpart.

Here, we will calculate the properties of this steady
state using a recently established variational principle
\cite{Weimer2015}.  In a spirit similar to equilibrium thermodynamics,
where a free energy functional has to be minimized, we consider a 
functional for dissipative systems that becomes nonanalytic at
a dissipative phase transition. To be specific, we will choose our
variational manifold as product states of the form
\begin{equation}
  \rho = \prod\limits_i \rho_i,\;\;\;\;\;
 \rho_{i} =\frac{1}{2} \left( 1+ \sum\limits_{\mu\in \{x,y,z\}} \alpha_\mu^{\phantom \dagger} \sigma_\mu^{(i)} \right).
 \label{eq:varstate}
\end{equation}
Later on, we will investigate in detail the validity of this approach
by explicitly considering fluctuations around product states. In the
case of product states, the variational principle is based on the
minimization of $ D= \sum_{\langle ij \rangle }||\dot \rho_{ij}||$
\cite{Weimer2015}. Here, the norm $||\dot \rho_{ij}||$ is given by
the trace norm, $||x||=\text{Tr}\{|x|\}$, and $\dot \rho_{ij}$ is the
reduced two-site operator obtained after taking the partial trace of
the time derivative $d \rho /dt $, according to $\dot
\rho_{ij}=\text{Tr}_{\not \phantom i i \not j}\{d \rho /dt \}$. As our
model is translationally invariant, it is sufficient to consider the
variational norm of a single bond $||\dot \rho_{ij}||$. Then, the
steady state is approximated by the variational minimization procedure
$||\dot \rho_{ij}|| \rightarrow \text{min}$. We would like to stress
that although our ansatz according to Eq. (\ref{eq:varstate}) is a
product state, the variational principle differs from a pure
mean-field decoupling as $\dot \rho_{ij}$ includes the time derivative
of correlation functions \cite{Weimer2015}.

Next, we perform an expansion of the variational norm $||\dot \rho_{ij}||$ in the order parameter $\phi \equiv \langle \sigma_x \rangle$,
in close analogy to Landau theory for equilibrium phase transitions. The degree of non-analyticity of the order parameter can be used to classify the phase transition: a discontinuous jump indicates a first order transition, while a diverging derivative corresponds to a second order transition. Within our product state approach, we choose the variational parameters according to
\begin{equation}
\alpha=\left (\langle \sigma_x \rangle,\langle \sigma_y \rangle , \langle \sigma_z \rangle \right )=(\phi, c\phi , \lambda). 
\label{eq:alpha}
\end{equation}
Separating the order parameter $\phi$ in the $\langle \sigma_y \rangle $ expression
has the advantage that $c$ becomes an analytic function. In the
following, we will choose $\lambda$ such that we always have a pure
state satisfying $|\alpha|^2 = 1$. Taking $\lambda$ as an independent
variational parameter does not lead to a significant difference in our
results, i.e., solution close to phase boundaries exhibit high purity.

Expanding $||\dot \rho_{ij}||$ up to the sixth order in $\phi$ leads to
\begin{equation}
||\dot \rho_{ij}||= u_0+u_2 \phi^2+u_4 \phi^4+u_6 \phi^6
\label{eq:normexp}
\end{equation}
as odd powers in $\phi$ vanish because of the $Z_2$ symmetry. From
the exact diagonalization of the $4\times 4$ matrix $\dot{\rho}_{ij}$, we
can readily calculate the expansion coefficients $u_n$ as functions of
the coupling constants $J$, $\Delta$, and $\gamma$, as well as the
coordination number $z$ and the variational parameter $c$.

As the next step, we determine the variational solution for the
parameter $c$. According to our ansatz of Eq.~(\ref{eq:alpha}), the
non-analytic behavior is contained in $\phi$, whereas $c$ is a smooth
function.  Therefore the value of $c$ close to phase boundaries is
fixed by its behavior far away from phase transitions. In the latter
regime, $\phi^2$ is the leading order of the variational functional
Eq.~(\ref{eq:normexp}), which allows us to find the variational
minimum by minimizing only $u_2$. Doing so with respect to $c$ leads
to
\begin{align}
 c=\frac{J \gamma z}{(\gamma/2)^2+4\Delta^2}.
 \label{eq: c}
\end{align}
Using that expression for $c$, there is only the order parameter
$\phi$ left as an independent variational parameter. Consequently, we
have successfully constructed the equivalent of Landau theory for
dissipative phase transitions and determined all expansion parameters
from the microscopic quantum master equation \footnote{See
  Appendix A for the full expression of the expansion
  coefficients.}.

\begin{figure}[t!]
\includegraphics[width=1\linewidth]{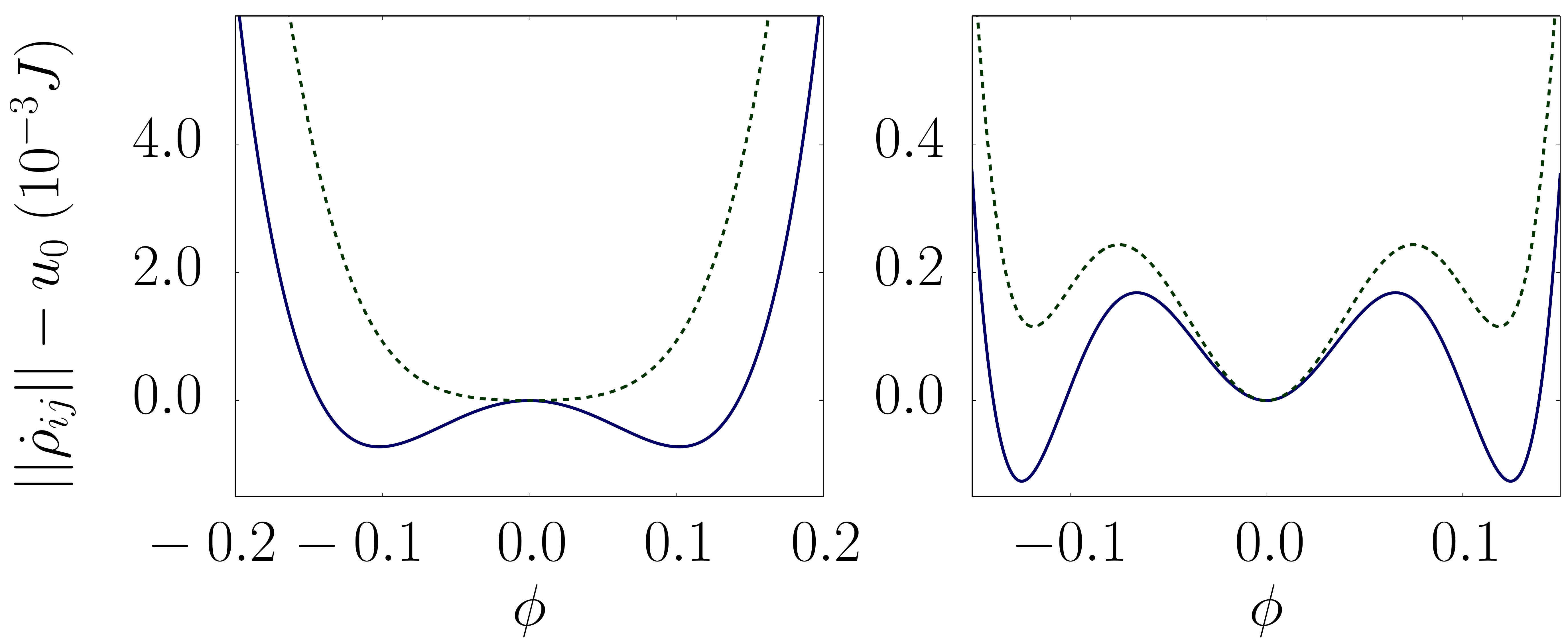}
\caption{Expansion in the variational norm $||\dot \rho_{ij}||$ to the sixth order in $\phi$ close to the second order transition (left) and 
close to the first order transition (right). In the ferromagnetic phase, the minimal variational norm is found at $\phi \ne 0$ (solid lines).
In the paramagnetic phase, the global minimum is located at $\phi=0$ (dashed lines).
}
\label{fig:varnorm}
\end{figure}
The dissipative functional of Eq.~(\ref{eq:normexp}) is mathematically
equivalent to the free energy functional of a $\phi^6$ theory, whose
possible phases are known \cite{Chaikin1995}. For $u_4>0$, the
$\phi^6$ term is irrelevant, and there is a continuous Ising
transition between a paramagnetic phase ($u_2 > 0$) and a
ferromagnetic phase ($u_2 < 0$). Close to the transition, the order
parameter behaves as $\phi = \pm (|u_2|/2u_4)^{1/2}$. In the
equilibrium Ising model, $u_4$ is always positive, but here we find
that this is not the case when adding dissipation. If the dissipation
rate $\gamma$ is sufficiently larger than the transverse field
$\Delta$, $u_4$ will become negative, which substantially alters the
phase diagram of the model. In order to find a stable variational
solution, it is then necessary to also consider the $\phi^6$ term of
the series expansion. We find that the variational norm has three
different minima, which transforms the transition between the
paramagnetic and the ferromagnetic phase into a first order
transition, see Fig.~\ref{fig:varnorm}. Remarkably, the $\phi^6$
theory exhibits a tricritical point at $u_2 = u_4 = 0$, which belongs
to a universality class different from that of the Ising
transition. This change of the universality class can be seen from the
scaling of the order parameter along the $u_4=0$ line, $\phi = \pm
(|u_2|/3u_6)^{1/4}$, which exhibits a different critical
exponent \footnote {We would like to stress that this situation is
  different from other cases of multicritical behavior in dissipative
  systems, i.e., where already the equilibrium model has a tricritical point
  \cite{Keeling2014}, or where there is no transition at all in the
  equilibrium model \cite{Marcuzzi2016}.}.
We have also confirmed the validity of our series
  expansion in $\phi$ by comparison to a numerical minimization of the
  variational norm including all orders. The full phase diagram of
the dissipative Ising model is shown in Fig.~1.

\emph{Fluctuations.---} So far, we have neglected the fact that the
true steady state of the system is not a product state. In reality,
there will be fluctuations in the system that lead to deviations from
the variational solution of the series expansion of
Eq.~(\ref{eq:normexp}). Importantly, the strength of these
fluctuations is inherently determined by the value of the variational
norm at the variational minimum. In close analogy to equilibrium
transitions, we can analyze at which point fluctuations lead to a
breakdown of the product state ansatz. To take these
  fluctuations into account, it is first necessary to introduce
  spatial inhomogeneities of the order parameter. Then, fluctuations generate such spatial inhomogeneities in the same manner as in equilibrium systems.
To this end, we will take
long-wavelength inhomogeneities into account by performing a gradient
expansion of the variational norm. Then we can evaluate the
equivalent of the Ginzburg criterion \cite{Kleinert2001} to determine
the range of validity of our effective theory.

We first allow for spatial variations within our product
state ansatz. Then the variational functional can be written as \footnote{See
  Appendix B for a detailed derivation.}
\begin{align}
 D= \sum\limits_{\langle ij \rangle }||\dot \rho_{ij}|| = & \sum_{\langle ij \rangle } z \left[\frac{J}{2}\left(1-\frac{1}{z}\right)+\frac{J'}{z} \right](\phi_i-\phi_j)^2  \notag
 \\ + & \sum_{ i } z \left [ u_0+u_2 \phi_i^2+u_4 \phi_i^4+u_6 \phi_i^6 \right ],
 \label{eq:ineqgrad}
\end{align}
where $\phi_i= \langle \sigma_x^{(i)} \rangle$ is the value of the order
parameter field at site $i$ and the coupling constant $J'$ is given by
\begin{align}
 J'=-\frac{J}{4}+\frac{\left (\frac{\gamma}{4} \right )^2 + \Delta^2}{4 J}+\frac{J \gamma^2}{\gamma^2+16 \Delta^2}. 
\end{align}
The first term in Eq.~(\ref{eq:ineqgrad}) describes spatial variations
of the order parameter to lowest order, while the other terms
correspond to the original series expansion of
Eq.~(\ref{eq:normexp}). For a finite value of
  $\phi_i$, the eigenbasis of $\rho_i$ is rotated away from the
  eigenbasis of $\sigma_z$. Consequently, the coupling constant $J'$
  also depends on $\gamma$ and $\Delta$. Taking the continuum limit,
we arrive at a Ginzburg-Landau-like functional for the variational
norm,
\begin{equation}
D[\Phi]= z\int d^{ d} x ~ u_0 + v_2 (\nabla \Phi)^2  + u_2 \Phi^2 + u_4 \Phi^4 + u_6 \Phi^6,
\end{equation}
where the order parameter $\phi$ follows from spatial averaging of the
fluctuating field $\Phi(x)$. The gradient term $v_2$ can then be readily identified as 
\begin{equation}
v_2 = \left[ \frac{J'}{z}+ \frac{J}{2}\left(1-\frac{1}{z}\right) \right]a^2,
\end{equation}
where $a$ is the lattice spacing, which we set to unity in the following.

Following from the existence of a dynamical symmetry
\cite{Sieberer2013}, fluctuations in the system will exhibit thermal
statistics at long wavelengths \cite{Maghrebi2016}. Hence, we can
characterize the strength of these fluctuations by an effective
temperature $T_\text{eff}$. Crucially, the strength of fluctuations is
determined by the value of the variational norm, as its value is a
measure of how much the exact steady state deviates from the product
state solution. However, we have to renormalize the variational norm
to get an intensive quantity. Then we find that the effective
temperature is connected to the variational norm according to
$T_{\text{eff}}=\frac{z}{2} || \dot \rho_{ij}||$, where the
variational norm $|| \dot \rho_{ij}||$ is to be evaluated in the
absence of spatial inhomogeneities, i.e., the choice of $i$ and $j$
does not matter. In the paramagnetic phase, the variational solution
results, according to the minimization of Eq. (\ref{eq:normexp}), in
$\alpha=(0,0,-1)$, which corresponds to a variational norm of $||\dot
\rho_{ij}||=2J$.  Remarkably, the resulting effective temperature on
the Ising transition line is given by $T_\text{eff} = zJ$, which
matches exactly the result found within the Keldysh formalism
\cite{Maghrebi2016}.

\begin{figure}[t!]
\centering
\includegraphics[width=1.0\linewidth]{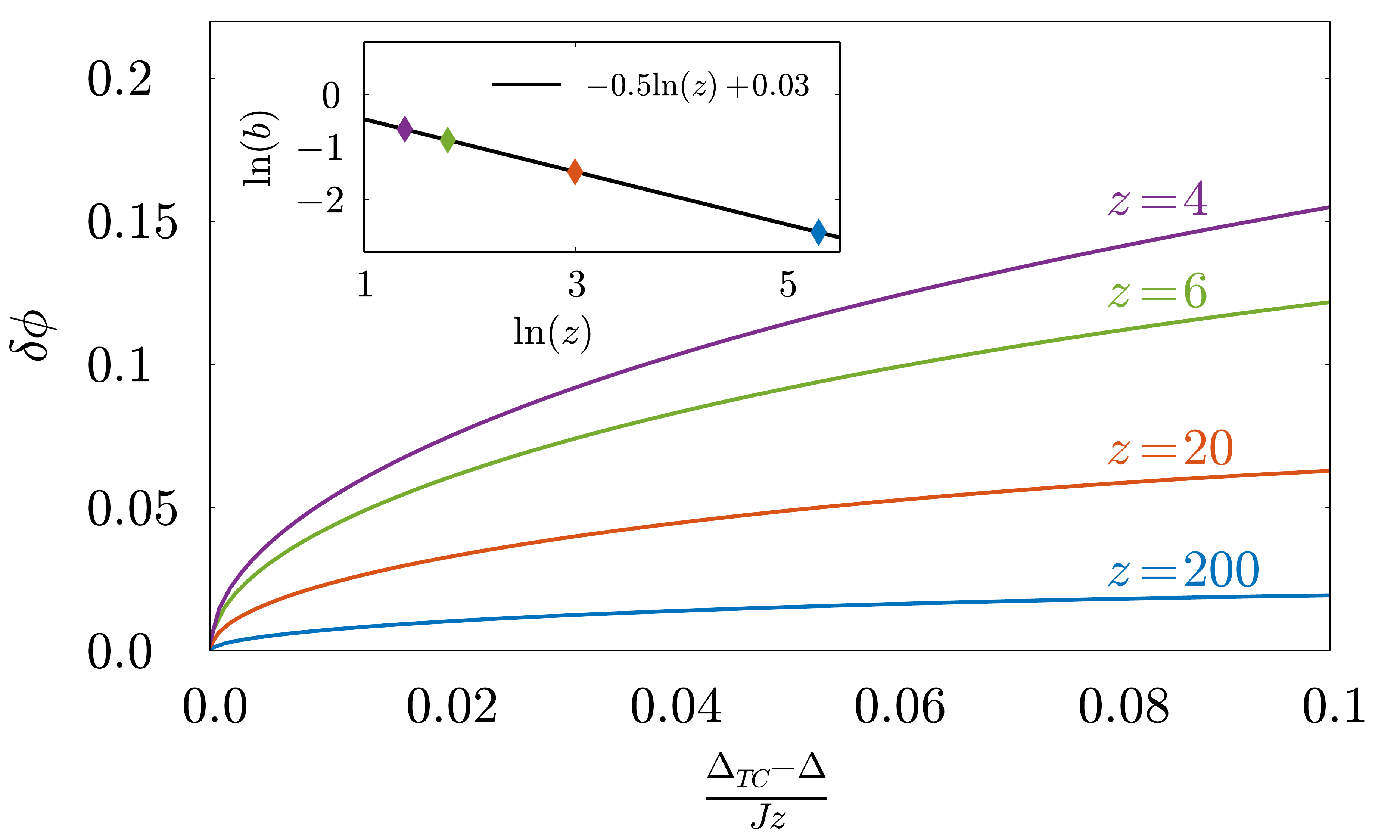}
\caption{First order jump $\delta \phi$ versus $\frac{\Delta_{TC}-\Delta}{Jz}$ along the first order line for $z=4,6,20$ and $200$. 
Inset: Logarithm of $b$, which is obtained from the fit of $\delta \phi$ according to Eq. (\ref{formula:deltaphi}), versus $\ln(z)$ and the corresponding fit (solid line).}
\label{fig:deltaepsloglog}
\end{figure} 

Using this effective temperature, we can now evaluate the strength of
fluctuations around the homogeneous solution. Considering Gaussian
fluctuations, we find for the mean squared fluctuations
\begin{align}
\langle [\phi-\Phi]^2 \rangle=\frac{T_{\text{eff}}}{2 v_2}\xi^{2-d} w^{d},
\label{formula:sqdeviation}
\end{align}
where $\xi^2=v_2/2|u_2|$ is the square of the correlation length, $d$
is the number of spatial dimensions, and $w = 0.0952 $ is a numerical
constant \cite{Kleinert2001}. Corresponding to the Ginzburg criterion,
we compare these mean squared fluctuations to the square of the order
parameter close to the multicritical point, which results in
\begin{align}
\frac{ \langle [\phi-\Phi]^2\rangle}{\phi^2} = \frac{\sqrt 3}{4} w^{d} v_2^{-d/2}u_0 \sqrt u_6  u_2^{(d-3)/2}.
\end{align}
The self-consistency of our effective theory is determined by the
exponent of the $u_2$ term. For $d>3$, the exponent is positive and
the relative strength of fluctuations is decreasing when approaching
the multicritical point, i.e., our effective theory becomes
self-consistent. For $d<3$, the exponent is negative and fluctuations
diverge close to the multicritical point. Hence, $d=3$ is the upper
critical dimension of the multicritical point, above which critical
exponents derived within Landau theory according to
Eq.~(\ref{eq:normexp}) become exact. At the experimentally accessible
case of $d=3$, one can expect merely logarithmic corrections to the
Landau theory exponents \cite{Kenna2004}.
We would like to point out that the same result can also be obtained from a renormalization group calculation, 
which also allows to evaluate corrections to the position of the tricritical point in a systematic way. 
While the position of the tricritical point is shifted significantly on including the renormalization group corrections in three dimensions,
we find that the strength of the shift decreases exponentially with increasing spatial dimensions \footnote{See Appendix C for a one-loop calculation based on the perturbative renormalization group.}.

\emph{Comparison to mean field results.---} In contrast to the
equilibrium case, mean field does not describe the correct physics at
the upper critical dimension in our open system as it misses the first
order transition and the tricritical point \cite{Maghrebi2016}.
Still, mean-field theory becomes exact as $d \rightarrow \infty$, where the variational approach and mean-field theory
agree. We now investigate in detail how the variational solution behaves as the dimensionality is increased. Specifically, we consider the value of the jump of the order parameter at the first order transition, $\delta\phi$, which is given by
\begin{align}
  \delta \phi=b\left (\frac{\Delta_{TC}-\Delta}{Jz} \right )^{1/2},
  \label{formula:deltaphi}
\end{align}
where $\Delta_{TC}$ is the value of $\Delta$ at the tricritical
point. Remarkably, the tricritical point remains at a finite value of
$\Delta$ even when the dimensionality of the system diverges,
asymptotically approaching $(\Delta/zJ,\gamma/zJ)_{TC}=(0.22,1.66)$ in
the limit of infinite spatial dimensions. Consequently, the mean-field
result is not recovered in a way that leads to a disappearance of the
tricritical point. Instead, the prefactor $b$ decreases according to
$b\sim 1/\sqrt{d}$ as the dimensionality of the system is increased,
see Fig.~\ref{fig:deltaepsloglog}. Hence, for any finite dimension,
the tricritical point can be observed and the mean-field prediction is
incorrect. Therefore, our results present further evidence (see also
\cite{Weimer2015a,Maghrebi2016}) that, for dissipative systems,
mean-field theory can be qualitatively incorrect even above the upper
critical dimension. Instead, it appears that only the variational
principle is capable of correctly describing this regime of high
dimensionality.  Finally, we find that, according to our variational
analysis, the location of the first order transition at the $\Delta=0$
line approaches the value $\gamma =0$ with increasing dimension. This
behavior is consistent with analytic arguments showing that there is
no ferromagnetic phase at $\Delta=0$ in any dimension
\footnote{A. V. Gorshkov et al., in preparation}.
\begin{figure}[t!]
\centering
\includegraphics[width=0.8\linewidth]{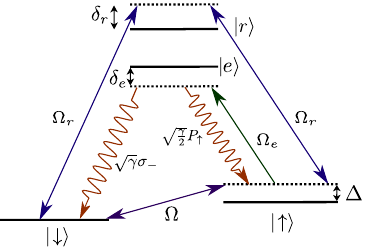}
\caption{4-level scheme with the ground states corresponding to the two spin configurations $| \! \uparrow \rangle$ and $| \! \downarrow \rangle$, and the dressed Rydberg state $|r\rangle$. 
The dissipation is realized via the $|e\rangle$-state.}
\label{fig:levelscheme}
\end{figure}

\emph{Experimental realization.---} For an experimental implementation
of the dissipative Ising model, we turn to a scenario where a Rydberg
state is weakly admixed to the electronic ground state manifold
\cite{Henkel2010,Pupillo2010,Honer2010,Glaetzle2015,vanBijnen2015}. Such Rydberg dressing
of ground state atoms has recently been observed in several
experiments \cite{Jau2015,Zeiher2016,Helmrich2016}. Here, we consider the dressing performed within
a Raman scheme, where two ground states are coupled to the same
Rydberg state, see Fig.~\ref{fig:levelscheme}. We obtain the
effective Hamiltonian for the dressed system based on fourth order
degenerate perturbation theory \cite{Lindgren1974,Fresard2012} in
$\Omega_r/\delta_r$, which generically has the form
\begin{equation}
  H = \Delta \sum\limits_i \sigma_z^{(i)} + \Omega' \sum\limits_i \sigma_x^{(i)} - \sum\limits_{ij} J_{ij} \sigma_x^{(i)}\sigma_x^{(j)}~+~\text{const}.
\end{equation}
This Hamiltonian is not yet in a $Z_2$ symmetric form as the
$\Omega'$ term breaks the symmetry. Crucially, this symmetry-breaking
term can be canceled by including a direct coupling $\Omega$ between
the two ground states into our perturbative analysis, see
Fig.~\ref{fig:levelscheme}.
Choosing $\Omega_r = \delta_r/10$ and $\Delta\sim |J_{\langle
ij\rangle}| \sim \Omega_r^4/\delta_r^3$ allows one to suppress the
strength of all $Z_2$ symmetry-breaking terms by several
orders of magnitude. Tuning the Rydberg interaction strength $V$ such
that $V=3\,\delta_r$ ensures that the effective interaction potential
can be cut off beyond nearest neighbors.

Finally, we realize the dissipative terms by performing optical
pumping from the spin-up into the spin-down state. In the case of
$^{87}\text{Rb}$, this can be realized by choosing
$\mathopen{|}\uparrow\mathclose{\rangle} = | 5S_{1/2}, F=2, m_F=2
\rangle$, $\mathopen{|}\downarrow\mathclose{\rangle} = |5S_{1/2}, F=2,
m_F=1 \rangle$, and $|e\rangle = |5P_{3/2}, F=3, m_F=2 \rangle$. Note
that this will result in an additional dephasing term described by the
jump operator $P_{\uparrow} =
\mathopen{|}\uparrow\mathclose{\rangle}\hspace{-0.25em}\mathopen{\langle}
\uparrow\mathclose{|}$, however, this term preserves the $Z_2$
symmetry and is also weaker than the dissipative spin flip
\cite{Steck2001a}. Finally, we would like to mention
  that the dissipative Ising model can also be realized within the
  experimental implementation suggested in \cite{Lee2013}, at the
  expense of requiring additional laser fields.

\begin{acknowledgments}

  We acknowledge fruitful discussions with T.~Vekua. This work was
  funded by the Volkswagen Foundation and the DFG within RTG 1729 and
  SFB 1227 (DQ-mat). M.F.M. and A.V.G acknowledge funding from NSF
  QIS, AFOSR, ARL CDQI, NSF PFC at JQI, ARO, and ARO MURI.

\end{acknowledgments}

\appendix
\setcounter{secnumdepth}{0}

\section{Appendix A: Expansion coefficients of the variational norm}
The expansion coefficients of the variational norm according to
\begin{equation}
||\dot \rho_{ij}||= u_0+u_2 \phi^2+u_4 \phi^4+u_6 \phi^6
\label{eq:normexp}
\end{equation}
are given by

\onecolumngrid

\begin{align}
  u_0 &=  2J, ~~~
 u_2 = \frac{\frac{\gamma ^2}{16}+\Delta ^2}{J}+J \left(\frac{16 \Delta ^2 z^2}{\gamma ^2+16 \Delta ^2}-1\right)-2 \Delta  z  ,    
\\
 u_4 &=  -\frac{1}{512 J^3 \left(\gamma ^2+16 \Delta ^2\right)^4} \biggl[\left(\gamma ^2+16 \Delta ^2\right)^6+8192 \gamma ^5 J^7 z^4+131072 \gamma ^4 \Delta ^2 J^6 z^4 
 \\ \notag & -1024 \gamma ^2 J^5 z^2 \left(\gamma ^2+16 \Delta ^2\right)^2 (8 \Delta  z-\gamma )
  +16384
   \Delta ^2 J^4 z^2 \left(\gamma ^2+16 \Delta ^2\right)^2 \left(\gamma ^2+4 \Delta ^2 z^2\right)
   \\ \notag &
    +32 J^3 \left(\gamma ^2+16 \Delta ^2\right)^3 \left(\gamma ^3+16 \gamma  \Delta ^2+256 \Delta ^3 z
   \left(1-2 z^2\right)+16 \gamma ^2 \Delta  z\right)
   \\ \notag & 
   -64 J^2 \left(\gamma ^2+16 \Delta ^2\right)^4 \left(\gamma ^2+8 \Delta ^2 \left(1-3 z^2\right)\right)-64 \Delta  J z \left(\gamma ^2+16 \Delta
   ^2\right)^5   ],
\\
 u_6 = & -\frac{1}{24576
    J^5  (\gamma ^2+16 \Delta ^2  )^6}  \biggl[- (\gamma ^2+16 \Delta ^2  )^9+1048576 \gamma ^7 J^{11} z^6 
    -524288 \gamma ^6 J^{10} z^6  (\gamma ^2-16 \Delta ^2  )
     \\ \notag &
     -65536 \gamma ^4 J^9 z^4
    (\gamma ^2+16 \Delta ^2  )   (-3 \gamma ^3+16 \gamma  \Delta ^2  (2 z^2-3  )+8 \gamma ^2 \Delta  z+128 \Delta ^3 z    )
    \\ \notag &
    -131072 \gamma ^4 J^8 z^4  (\gamma ^2+16 \Delta
   ^2  )   (\gamma ^4-8 \gamma ^2 \Delta ^2+128 \Delta ^4  (2 z^2-3  )-2 \gamma ^3 \Delta  z-32 \gamma  \Delta ^3 z    )
    \\ \notag & 
    +4096 \gamma ^2 J^7 z^2  (\gamma ^2+16 \Delta ^2  )^2
   (\gamma ^5  (3-2 z^2  )-96 \gamma ^3 \Delta ^2  (z^2-1  )+1536 \gamma ^2 \Delta ^3 z  (z^2-1  )\\ \notag &
  +256 \gamma  \Delta ^4  (3-4 z^2  )+4096 \Delta ^5 z  (2
   z^2-3  )-48 \gamma ^4 \Delta  z   ) 
   -2048 J^6 z^2  (\gamma ^2+16 \Delta ^2  )^3   (5 \gamma ^6+64 \gamma ^4 \Delta ^2  (3 z^2-1  ) \\ \notag & +256 \gamma ^2 \Delta ^4  (14
   z^2-9  )
    +8192 \Delta ^6 z^2  (z^2-1 )-16 \gamma ^5 \Delta  z-256 \gamma ^3 \Delta ^3 z   ) \\ \notag & -256 J^5  (\gamma ^2+16 \Delta ^2 )^4   (\gamma ^5  (4 z^2-1 )-8
   \gamma ^4 \Delta  z  (4 z^2+3 )+32 \gamma ^3 \Delta ^2  (3 z^2-1 )
   \\ \notag &
   -256 \gamma ^2 \Delta ^3 z  (4 z^2+3 )+256 \gamma  \Delta ^4  (2 z^2-1 )-6144 \Delta ^5 z
    (1-2 z^2 )^2   ) \\ \notag & -256 J^4  (\gamma ^2+16 \Delta ^2 )^5   (5 \gamma ^4+16 \gamma ^2 \Delta ^2  (7-10 z^2 )
    +256 \Delta ^4  (15 z^4-12 z^2+2 )\\ \notag &
    -4 \gamma ^3 \Delta  z 
   -64 \gamma  \Delta ^3 z   )
   -32 J^3  (\gamma ^2+16 \Delta ^2 )^6   (\gamma ^3+16 \gamma  \Delta ^2+1280 \Delta ^3 z  (1-2 z^2 )+112 \gamma ^2 \Delta  z   )
    \\ \notag &
   +16 J^2  (\gamma ^2+16 \Delta ^2 )^7   (5 \gamma ^2+48 \Delta ^2  (1-5 z^2 )   )+96 \Delta  J z  (\gamma ^2+16 \Delta ^2 )^8 \biggr]. 
   \end{align}

   \twocolumngrid
\section{Appendix B: Derivation of the variational norm including spatial fluctuations}
In this section, we derive Eq. (9) in the main text.

In order to evaluate the consistency of our product state ansatz, we
allow spatial inhomgeneities of the order parameter field. Consequently,
the variational parameter $\phi_{i}=\langle \sigma^{(i)}_x \rangle$
has a different value at each site $i$ in our ansatz for the product
state density matrix $\rho$.
 
Expanding the variational norm $||\dot \rho_{ij}||$ with respect to
the order parameter, we get terms of the form $u_{2n} \phi_i^{2n}$, which
correspond to the terms of the expansion in the homogeneous case. Due
to the fluctuations of the order parameter field, we get additional
gradient terms of the form $v_2 (\phi_k-\phi_l)^2$ in the lowest
(quadratic) order, where $k$ and $l$ are neighbouring sites. Here, we
have a contribution to the variational norm $||\dot\rho_{ij}||$ from
the difference between sites $i$ and $j$ and from the gradient between
site $i$ and $j$ and their nearest surrounding sites $k$ and $l$,
respectively.

The variational functional can then be written as
\onecolumngrid

\begin{align}
 D=\sum_{\langle ij \rangle }||\dot \rho_{ij}|| = \sum_{\langle ijkl \rangle }  \frac{J}{2}\left(z-1\right) (\phi_i-\phi_k)^2+\frac{J}{2}\left(z-1\right) (\phi_j-\phi_l)^2+J'(\phi_i-\phi_j)^2  +  \sum_{\langle ij \rangle } \left [ u_0+u_2 \phi_i^2+u_4 \phi_i^4+u_6 \phi_i^6 \right ],
\end{align}
\twocolumngrid
with
\begin{align}
  J'=-\frac{J}{4}+\frac{\left (\frac{\gamma}{4} \right )^2 + \Delta^2}{4 J}+\frac{J \gamma^2}{\gamma^2+16 \Delta^2}.
\end{align}
In the long wavelength limit, we have $\phi_i-\phi_k = \phi_j-\phi_l = \phi_i-\phi_j$, and after factoring out the coordination number $z$ we arrive at
\begin{align}
D= & \sum_{\langle ij \rangle } z \left[\frac{J}{2}\left(1-\frac{1}{z}\right)+\frac{J'}{z} \right](\phi_i-\phi_j)^2 \\ & \notag +
  \sum_{ i } z \left [ u_0+u_2 \phi_i^2+u_4 \phi_i^4+u_6 \phi_i^6 \right ].
\end{align}
\section{Appendix C: Renormalization group correction of the tricritical point}

\begin{figure}[h!]
\includegraphics[width=0.75\linewidth]{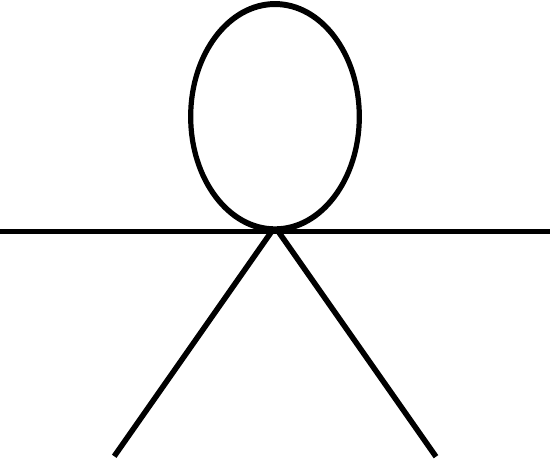}
\caption{Diagrammatic visualization of the one-loop correction of the $u_4$-term. The branches represent the order parameter field in fourth order, the circle stands for the contracted part of the momentum space integral.
}
\label{fig:loopdiag}
\end{figure}

In the following, we will calculate the shift of the tricritical point when renormalization group corrections of the $u_4$ term are included.
Starting with the Ginzburg-Landau functional 
\begin{equation}
D[\Phi]= z\int d^{ d} x ~ u_0 + v_2 (\nabla \Phi)^2  + u_2 \Phi^2 + u_4 \Phi^4 + u_6 \Phi^6,
\label{glfunc}
\end{equation}
a perturbative momentum space renormalization group analysis leads to the linear flow equations \cite{Wilson1974}
\begin{align}
 \frac{du_2}{dl} & =2 u_2+c_1 u_4+c_2 u_6 \\
 \frac{du_4}{dl} & =(4-d)u_4+c_3 u_6 \\ 
 \frac{du_6}{dl} & =(3-d) u_6.
 \label{eq:flow}
\end{align}
Here, the $c_i$ are constants that follow from the one-loop expansion of the interaction terms. In particular, the $c_3 u_6$- term stems from the one loop diagram shown in Fig.~\ref{fig:loopdiag}. The value of the $c_3$-coefficient is given by 
\begin{align}
 c_3=\frac{2^{-d} 15 S_d}{\pi^2 v_2},
\end{align} 
where $S_d$ is the surface area of the $d$-dimensional unit sphere. 
Here, we made a cutoff of the momentum space integral at $\Lambda=\pi/a$, where $a=1$ is the lattice spacing.

In the following, we choose $u_2 (0)$ such that we arrive at the
fixed point $u^{\ast}_2$ corresponding to the Ising critical line. Then,
the solution to the second equation will tell us about the nature of
the transition \cite{Goldenfeld1992}. For $u^\ast_4 = \infty$, we have
the conventional Ising transition, as the renormalized $u_4$ is
positive. For $u^\ast_4 = -\infty$, we get the first order transition,
while $u^\ast_4 = 0$ is the tricritical point (in $d \ge
3$). Depending on the initial values $u_4 (0)$ and $u_6 (0)$, we may
end up in any of these fixed points, allowing us to relate the
microscopic coupling constants $u_4 (0)$ and $u_6 (0)$ to the nature
of the transition and hence to the position of the tricritical
point. Using $\epsilon=3-d$, we arrive at the solutions
\begin{align}
 u_4(l) & =u_4(0) e^{(\epsilon +1)l}+c_3 u_6(0) \left[e^{(\epsilon +1)l}-e^{\epsilon l} \right ] \\
 u_6(l) & =u_6(0) e^{\epsilon l}
\end{align}
with $\epsilon=3-d$.  From the first equation, we can immediately see
that the sign of the fixed point depends on the sign of $u_4 (0)$ + $c_3
u_6 (0)$.  Hence, the position of the tricritical point is shifted
from $u_4 = 0$ in Landau theory to $u_4 = - c_3 u_6$ by the one-loop
correction. For $d=3$, we find that the shifted tricritical point is
located at $(\Delta/Jz,\gamma/Jz)_{\text{TC}}=(0.023,
0.35)$. In higher dimensions, the deviation from the variational solution
of the tricritical point decreases exponentially with the number
of spatial dimensions.

\bibliographystyle{aip}
\bibliography{../../Papers/bib/bib}

\end{document}